\documentclass[conference]{IEEEtran}
\IEEEoverridecommandlockouts
\usepackage{cite}
\usepackage{amsmath,amssymb,amsfonts}
\usepackage{graphicx}
\usepackage{textcomp}
\usepackage{xcolor}

\usepackage{algorithm}
\usepackage{algpseudocode}
\usepackage{algorithmicx}

\usepackage{multirow}
\usepackage{flushend}
\usepackage{microtype}
\usepackage{booktabs}

\def\BibTeX{{\rm B\kern-.05em{\sc i\kern-.025em b}\kern-.08em
    T\kern-.1667em\lower.7ex\hbox{E}\kern-.125emX}}
\begin{document}

\title{Fast Diffusion-GAN Model for Generating Symbolic Music with Emotions}

\author{
\IEEEauthorblockN{
Jincheng Zhang, Gy\"orgy Fazekas, Charalampos Saitis
}
\IEEEauthorblockA{
Centre for Digital Music, Queen Mary University of London, UK
}
}

\maketitle

\begin{abstract}
Diffusion models have shown promising results for a wide range of generative tasks with continuous data, such as image and audio synthesis. However, little progress has been made on using diffusion models to generate symbolic music since this class of generative models struggles with discrete data while its iterative sampling process is computationally expensive. In this work, we propose a diffusion model integrated with a Generative Adversarial Network, aiming to mitigate the slow sampling drawback of diffusion models applied to symbolic music generation. We enable the use of diffusion model by representing symbolic music as image-like pianorolls with velocity values instead of discrete event tokens and then used a Variational Autoencoder to obtain their latent representations to train the proposed diffusion model. Our results demonstrate the successful control of our diffusion model to generate symbolic music with a desired emotion. Our model uses merely 15 denoising diffusion steps to produce high-quality music, while most diffusion models for symbolic music generation require hundreds or thousands of steps.

\end{abstract}


\section{Introduction}
\label{sec:intro}

\begin{figure*}
\centering
\includegraphics[width=0.8\textwidth]{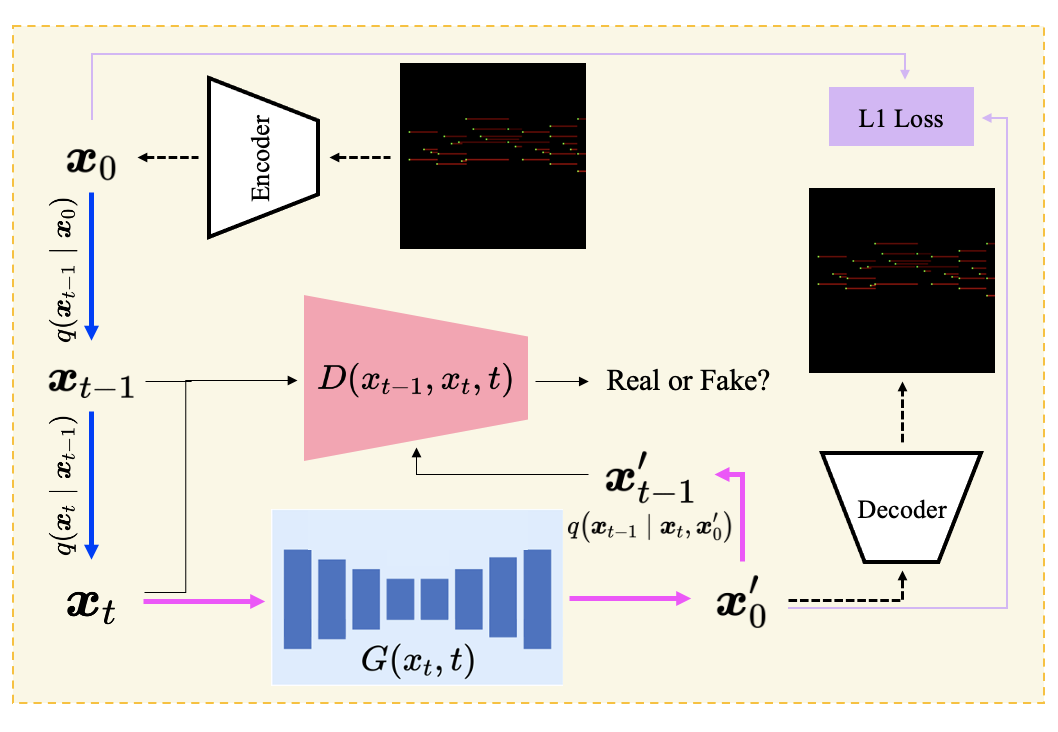}
\caption{\label{fig:MusicDiffusion}Illustration of our fast diffusion GAN model for discrete symbolic music generation. A GAN is used as our denoising network to generate new latent representations of pianorolls $\boldsymbol{x}_0^{\prime}$ which can be subsequently decoded back to symbolic music using the trained VAE. }
\end{figure*}

With the renaissance of artificial neural networks, the recent decade has witnessed the success of deep learning for numerous tasks including image processing \cite{yasrab2021predicting}, natural language processing \cite{li2022diffusion} and speech recognition \cite{kim2022squeezeformer}. Deep learning has also seen increasing use in symbolic music generation \cite{dong2018musegan}. However, good control of generative models to produce music with an anticipated goal remains challenging \cite{ferreira2021learning}. Without the satisfactory ability of control, the personalized requirements from different users may not be met, hindering the practical applications of those generative models. Compared to the generation of random music, controllable music generation can better facilitate the real-world applications of generative music systems because it allows the users to specify the desired musical attributes according to their own preferences and intents. Diverse users such as artists, music composers and filmmakers will gain significant benefits if music generation can be controlled. For instance, controllable generation systems can help filmmakers produce appropriate background music that is a good fit for a specific film scene. This paper presents a deep learning-based model and related experiments for symbolic music generation controlled by emotions. 

Few deep learning-based algorithms allow users to effectively specify the emotion of the generated music. Long short-term memory (LSTM) networks have been applied to compose symbolic music with a given emotion in terms of positive or negative valence \cite{ferreira2021learning}. However, LSTMs have become less popular because of their weaker capability of modeling long-term dependencies compared to Transformers \cite{vaswani2017attention}. Hung et al. \cite{hung2021emopia} proposed a ``compound'' word representation (CP) \cite{hsiao2021compound} and a Transformer-based model to generate symbolic music conditioned on categorical emotions. The CP represents symbolic music as discrete tokens, which are not inherently well-suited for diffusion models. Music FaderNets \cite{tan2020music} is a music style transfer model that adjusts the arousal of existing symbolic music instead of generating a new piece from scratch. 

Each of the two main classes of generative models employed in symbolic music generation previously, namely Generative Adversarial Networks (GANs) \cite{goodfellow2020generative} or Variational AutoEncoders (VAEs) \cite{kingma2013auto}, has its own trade-off between sample quality and mode coverage. GANs can synthesise high-quality data, but their generator module often learns to fool its discriminator module by generating samples with limited diversity, resulting in the well-known mode collapse effects \cite{xiao2021tackling}. Conversely, while VAEs cover the underlying data distribution better, they often suffer from low sample quality. In continuous data domains, diffusion models \cite{ho2020denoising} have
emerged as powerful alternatives as they can produce samples of state-of-the-art quality, higher distribution coverage, and a more stable training objective than GANs \cite{dhariwal2021diffusion}. However, the success of diffusion models has not been fully extended to the controllable generation of symbolic music, which is typically represented as
sequences of discrete tokens.

The applications of diffusion models to discrete data remain restricted. Only a few prior works have investigated the use of diffusion models for symbolic music generation. MelodyDiffusion \cite{li2023melodydiffusion}, using a transformer-based diffusion model, is limited to the generation of monophonic melodies. Zhang et al. \cite{zhang2024composer} proposed a discrete diffusion model combined with a VQVAE to generate symbolic music with target composer styles, where composer embeddings are injected into the denoising network through Adaptive LayerNorm (AdaLN) \cite{ba2016layer}. By contrast, this study employs classifier-free guidance \cite{ho2022classifier} to enable music generation conditioned on emotions.  Proffusion \cite{zhang2025mamba} is a diffusion model that incorporates a Mamba block and learnable wavelet transform for producing symbolic music controlled by chords. However, it needs 1000 denoising diffusion steps and thus results in the disadvantage of slow sampling. The closest work to ours is \cite{mittal2021symbolic} where Mittal et al. used diffusion models for infilling and unconditional generation of symbolic music by training diffusion models on symbolic music's continuous embeddings produced by a pre-trained MusicVAE \cite{roberts2018hierarchical}. The distinct differences are that our proposed diffusion model takes much fewer time steps to generate symbolic music by leveraging GANs. Moreover, our proposed diffusion model offers controls to produce symbolic music with specified emotions.

Xiao et al. \cite{xiao2021tackling} proposed an unconditional diffusion GAN for continuous image synthesis in a pixel space. By contrast, we trained our diffusion model on the symbolic music’s continuous embeddings produced by a trained VAE, allowing the use of diffusion model for the generation of symbolic music often represented as sequences of discrete event tokens. Additionally, our model uses classifier-free guidance for symbolic music generation conditioned on emotion which is one of the most important music attributes \cite{barthet2013music}. In summary, our main contributions are as follows: 
\begin{itemize}
\item In this work, we enable the use of diffusion models for generating symbolic music by transforming MIDI to pianorolls and then obtaining their continuous latent representations. While pianorolls are typically binary and lack performance features, our work extends this by incorporating velocity information into pianorolls, allowing the generation of expressive piano performances. 
\item The proposed model leverages GANs to drastically accelerate the sampling process of diffusion models for fast production of symbolic music.
\item We incorporate a L1 loss to complement the adversarial loss, as the former provides the generator with direct supervision and thus leads to more efficient training and improved convergence.
\item Our proposed model is capable of providing users with controls to generate symbolic music with target emotions.
\end{itemize}




\section{Method}

\subsection{Variational Autoencoder}

We design the VAE architecture following VQGAN \cite{esser2021taming}. Since prior work \cite{rombach2022high} has demonstrated the benefits of using two-dimensional latent variables over common one-dimensional vectors in the image compression and reconstruction domains, two-dimensional latent representations are used as both the encoder output and the input to the diffusion model stage.

\subsection{Latent Music Diffusion GAN Model}
\label{ssec:network}

Standard diffusion models include a forward process and a reverse process. In the forward diffusion process, Gaussian noise is progressively added to the original data $\mathbf{x}_{0}$ in $\boldsymbol{T}$ diffusion steps until the result approximates pure Gaussian noise, leading to a sequence of noisy samples $\mathbf{x}_{1}, \ldots, \mathbf{x}_{T}$ with the same dimensionality as the data $\mathbf{x}_{0}$:
\begin{equation}
q\left(\mathbf{x}_{t} \mid \mathbf{x}_{t-1}\right)=\mathcal{N}\left(\mathbf{x}_{t} ; \sqrt{1-\beta_{t}} \mathbf{x}_{t-1}, \beta_{t} \mathbf{I}\right)
\end{equation}
where $\beta_{1}, \ldots, \beta_{T}$ is the pre-defined variance schedule that controls the amount of noise added at each diffusion step. The posterior probability of the forward process $q\left(\mathbf{x}_{1: T} \mid \mathbf{x}_{0}\right)$ contains no trainable parameters. 

\begin{equation}
q\left(\mathbf{x}_{1: T} \mid \mathbf{x}_{0}\right)=\prod_{t=1}^{T} q\left(\mathbf{x}_{t} \mid \mathbf{x}_{t-1}\right)
\end{equation}


In the reverse process, a neural network such as a U-Net or a Transformer is used to learn the conditioned probability distributions $p_{\theta}\left(\mathbf{x}_{t-1} \mid \mathbf{x}_{t}\right)=\mathcal{N}\left(\mathbf{x}_{t-1} ; \boldsymbol{\mu}_{\theta}\left(\mathbf{x}_{t}, t\right), \boldsymbol{\Sigma}_{\theta}\left(\mathbf{x}_{t}, t\right)\right)$. Gaussian noise $\mathbf{x}_{T} \sim \mathcal{N}(\mathbf{0}, \mathbf{I})$ is iteratively denoised to approximate samples from the target data distribution: 
\begin{equation}
p_{\theta}\left(\mathbf{x}_{0: T}\right)=p\left(\mathbf{x}_{T}\right) \prod_{t=1}^{T} p_{\theta}\left(\mathbf{x}_{t-1} \mid \mathbf{x}_{t}\right)
\end{equation}

Equation \eqref{eq:6} defines the training objective where parameters $\theta$ can be learned by minimizing the negative log-likelihood of $p_\theta\left(\mathbf{x}_0\right)=\int p_\theta\left(\mathbf{x}_{0: T}\right) d \mathbf{x}_{1: T}$ (see \cite{ho2020denoising} for derivation details):


\begin{equation} \label{eq:6}
\begin{split}
&\mathbb{E}_{q}[{D_{\mathrm{KL}}\left(q\left(\mathbf{x}_{T} \mid \mathbf{x}_{0}\right) \| p_{\theta}\left(\mathbf{x}_{T}\right)\right)}
{-\log p_{\theta}\left(\mathbf{x}_{0} \mid \mathbf{x}_{1}\right)}\\
+&\sum_{t>1} {D_{\mathrm{KL}}\left(q\left(\mathbf{x}_{t-1} \mid \mathbf{x}_{t}, \mathbf{x}_{0}\right) \| p_{\theta}\left(\mathbf{x}_{t-1} \mid \mathbf{x}_{t}\right)\right)}]
\end{split}
\end{equation}

We propose a music generative model based on latent diffusion models and GANs, as shown in Fig. \ref{fig:MusicDiffusion}. This method begins with a trained encoder that compresses the input data $X_{0}$, mapping it from pixel space to a lower-dimensional latent representation $x_{0}=\mathcal{E}\left(X_{0}\right)$. Both the forward diffusion and reverse denoising processes in our model are performed in a low-dimensional latent space rather than the pixel space adopted by DDGAN \cite{xiao2021tackling}. This latent space strategy can significantly reduce the size of the input images, resulting in faster inference and lower computational cost. Furthermore, as demonstrated by Rombach et al. \cite{rombach2022high}, low-dimensional latent space is more suited than pixel space for diffusion models. This enables the model to generate samples with higher quality and greater diversity. 

Diffusion models commonly assume Gaussian distributions can be used to approximate the denoising distribution. However, the Gaussian assumption is justified only when the denoising step size is small \cite{sohl2015deep}, leading to the requirement of thousands of steps in the reverse process and thus the diffusion models' slow sampling issue. To enable a large step size, we model the denoising distribution using a multimodal GAN, as the former becomes more complex and multimodal when the number of denoising steps is reduced. Our forward diffusion is set up similarly to standard diffusion models, except that $T$ is small and each diffusion step has larger $\beta_{T}$.

Instead of directly predicting $\mathbf{x}_{t-1}^{\prime}$, a conditional GAN's generator is used to predict $\mathbf{x}_{0}^{\prime}$ before using the posterior distribution $q\left(\mathbf{x}_{t-1} \mid \mathbf{x}_{t}, \mathbf{x}_{0}^{\prime}\right)$ to sample $\mathbf{x}_{t-1}^{\prime}$ given $\mathbf{x}_{t}$ and the predicted $\mathbf{x}_{0}^{\prime}$. The sampling process can be expressed as Algorithm \ref{alg:lddgan_sampling}. Through adversarial learning, the time-dependent discriminator $D_\phi\left(\mathbf{x}_{t-1}, \mathbf{x}_t, t\right)$ will be able to discriminate whether $\mathbf{x}_{t-1}^{\prime}$ is a plausible denoised version of $\mathbf{x}_{t}$. Our discriminator is optimised by Eq. \eqref{eq:7} where fake samples from $p_\theta\left(\mathbf{x}_{t-1} \mid \mathbf{x}_t\right)$ are contrasted against real samples from $q\left(\mathbf{x}_{t-1} \mid \mathbf{x}_t\right)$.
The adversarial loss for the generator is defined as Eq. \eqref{eq:8}.
\begin{align} \label{eq:7}
\mathcal{L}_{a d v}^{D} & =-\log \left(D\left(x_{t-1}, x_{t}, t\right)\right)+\log \left(D\left(x_{t-1}^{\prime}, x_{t}, t\right)\right)  
\end{align}
\begin{align} \label{eq:8}
\mathcal{L}_{a d v}^{G} & =-\log \left(D\left(x_{t-1}^{\prime}, x_{t}, t\right)\right) 
\end{align}
In WDDGAN \cite{phung2023wavelet}, the training of the generator incorporates not only the adversarial objective but also a reconstruction loss between the generated sample and its ground truth. The inclusion of the reconstruction term has been shown to improve the fidelity of generated outputs \cite{phung2023wavelet}. This improvement stems from the fact that relying solely on adversarial training provides indirect guidance through the discriminator’s feedback, encouraging the generator to approximate the input $\mathbf{x}_{0}$ implicitly. In contrast, the reconstruction loss offers the generator direct supervision by comparing $\mathbf{x}_{0}$ and $\mathbf{x}_{0}^{\prime}$ using an L1 loss. As a consequence, combining reconstruction loss with adversarial loss leads to more efficient training and improved convergence. Therefore, a reconstruction loss is also adopted in this work. The overall loss for the generator is defined as a weighted sum of adversarial and reconstruction terms:
\begin{align}
\mathcal{L}_{\text {rec }}^{G} & =\left\|\mathbf{x}_{0}-\mathbf{x}_{0}^{\prime}\right\|  
\end{align}
\begin{align}
\mathcal{L}_{G} & =\mathcal{L}_{\text {adv }}^{G}+\lambda \mathcal{L}_{\text {rec }}^{G} 
\end{align}
where $\lambda$ is a weighting hyper-parameter.


\begin{center}

\begin{minipage}{0.80\columnwidth}  

\begin{algorithm}[H]
\caption{Sampling Process}
\label{alg:lddgan_sampling}
\begin{algorithmic}[1]

  \State $\mathbf{x}_T \sim \mathcal{N}(0, \mathbf{I})$
  \For{$t = T, \ldots, 1$}
    \State $\mathbf{z} \sim \mathcal{N}(0, \mathbf{I})$
    \State $\mathbf{x}'_{0} = G(\mathbf{x}_t, \mathbf{z}, t)$
    \State $\mathbf{x}'_{t-1} \sim q\left(\mathbf{x}_{t-1} \mid \mathbf{x}_{t}, \mathbf{x}_{0}^{\prime}\right)$
  \EndFor
  \State $\mathbf{X}_0 = \mathcal{D}(\mathbf{x}'_0)$
  \State \textbf{return} $\mathbf{X}_0$

\end{algorithmic}
\end{algorithm}
\end{minipage}
\end{center}

\begin{figure*}
\centering
\includegraphics[width=1.0\textwidth]{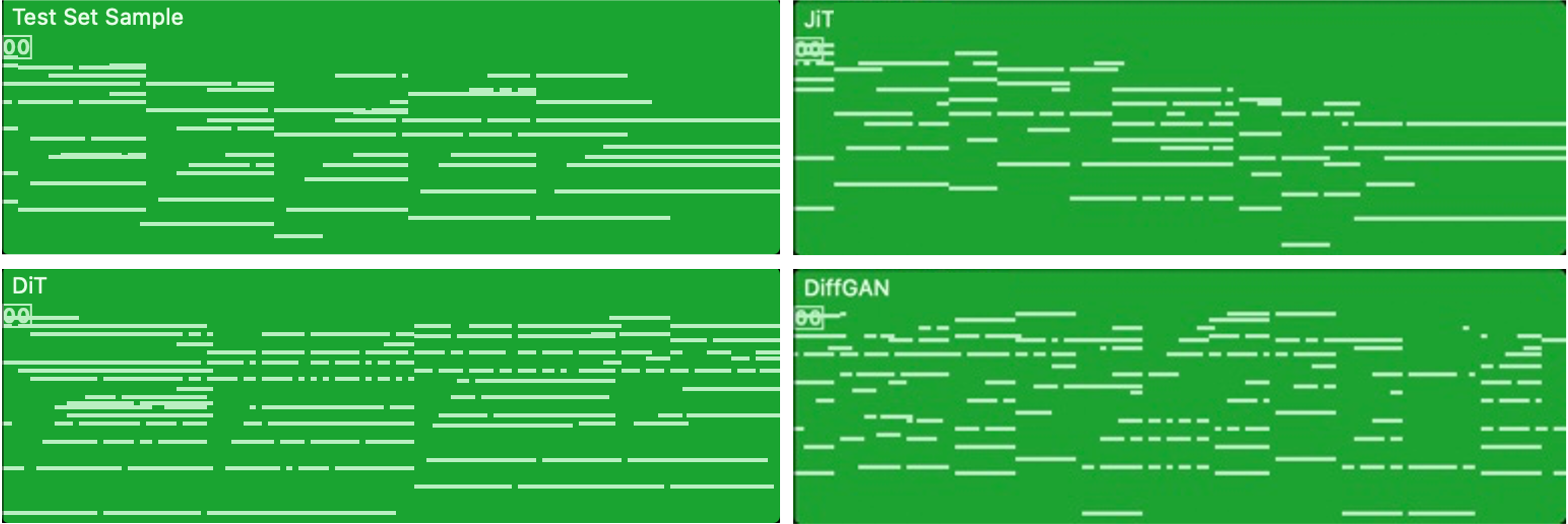}

\caption{\label{fig:music samples}Pianorolls generated by our fast DiffGAN, DiT and JiT, with pitch on the vertical axis and
time on the horizontal axis.}
\end{figure*}

Classifier guidance \cite{dhariwal2021diffusion} allows diffusion models to generate samples conditioned on $\mathbf{y}$ by training a classifier $p_{t}\left(\mathbf{y} \mid \mathbf{x}_{t}\right)$ on noisy samples paired with their corresponding conditions. During inference, the generation process is guided by combining the gradient of the log likelihood of this classifier with the diffusion model's score estimate:
\begin{equation}
\nabla_{\mathbf{x}_{t}} \log p\left(\mathbf{x}_{t}\right)+\omega \nabla_{\mathbf{x}_{t}} \log p_{t}\left(\mathbf{y} \mid \mathbf{x}_{t}\right)
\end{equation}
where $\omega$ denotes the guidance scale, determining the classifier guidance's strength. However, this approach necessitates training an additional classifier on noisy data. Therefore, we adopt classifier-free guidance \cite{ho2022classifier} method in this work, which eliminates the need for a separate classifier by jointly training a conditional and an unconditional diffusion model. During sampling, their score estimates are mixed as follows: 
\begin{equation}
(1+\omega) \nabla_{\mathbf{x}_{t}} \log p_{t}\left(\mathbf{x}_{t} \mid \mathbf{y}\right)-\omega \nabla_{\mathbf{x}_{t}} \log p_{t}\left(\mathbf{x}_{t}\right)
\end{equation}
The emotion labels are first embedded to continuous vectors using embedding layers. These condition vectors are then fed into both the generator and the discriminator. Classifier-free guidance is employed by randomly dropping the conditioning with a probability of 0.2 during training. A guidance scale of 1.1 is used during inference to generate pianoroll samples' latent representations which are subsequently decoded back to symbolic music.

\section{Experiments}
\subsection{Dataset}
\label{ssec:dataset}

Controllable music generation was investigated using the multi-modal EMOPIA dataset \cite{hung2021emopia}. It contains audio data and transcribed MIDI files of 1,087 pop piano music clips extracted from 387 songs, along with human-annotated emotion labels corresponding to the four quadrants of the commonly used Russell’s circumplex model of affect. This circular structure involves the two dimensions of arousal and valence, where valence denotes positive versus negative emotion and arousal indicates emotional intensity \cite{barthet2013music}. Specifically, the four classes of labels are: HVHA (high valence high arousal), LVHA (low valence high arousal), LVLA (low valence low arousal), and HVLA (high valence low arousal). To facilitate the use of diffusion models, the discrete MIDI files are transformed to pianorolls which resemble image formats with time on the horizontal axis and pitch on the vertical axis. Subsequently, a trained VAE is used to produce the pianorolls' latent representations for our diffusion model training.

\subsection{Experimental Setup}
\label{ssec:experiment}


Our fast music diffusion model uses Adam optimiser with $\beta_1$ = 0.5 and $\beta_2$ = 0.9. Initial learning rates for the generator and discriminator start from 1e-4 and 1.6e-4, respectively. Cosine learning rate decay is used to train both the generator and discriminator. The batch size is set to 512. We denote our proposed model as DiffGAN.

\subsection{Results \& Discussion}
\label{sec:print}



\begin{table}
\caption{Objective evaluation of music quality and emotion correspondence for the samples generated by different models. The emotion classifier's accuracy on the test set is 0.654.}
\centering
\renewcommand{\arraystretch}{1.35}
\resizebox{0.48\textwidth}{!}
{
\begin{tabular}{cccccc}
\specialrule{0.8pt}{0pt}{0pt}
 
\multirow{2}{*}{Model} & \multirow{2}{*}{Steps} & \multicolumn{2}{c}{Music Quality} & \multirow{2}{*}{Accuracy$\uparrow$}             \\ \cline{3-4}
& & PCHE$\rightarrow$ & GPS$\rightarrow$ &  \\ \hline
 
Real Data   & --                & 2.646                                  & 0.579              & 0.654             \\

DiT    & 250               &  2.803                                  & 0.604            & 0.510             \\
JiT    & 10               &  2.543                                  & 0.611            & 0.615             \\
DiffGAN &  5       & 2.779                         &  0.590        & 0.579    \\
DiffGAN &  10       & 2.729                         &  0.588        & 0.622    \\
DiffGAN &  15       & 2.707                         &  0.587        & 0.635    \\
\specialrule{0.8pt}{0pt}{0pt}
\end{tabular} 
}

\label{tab:model_comparison}
\end{table}

We compared our DiffGAN with a Diffusion Transformer (DiT) \cite{peebles2023scalable}, which operates in the same latent space as our model, as well as the recently proposed JiT \cite{li2025back}, which is a pixel-space diffusion model that also enables fast sampling. Additionally, we also examined how performance varies with the number of denoising diffusion steps. Each trained generative model was used to generate 1000 samples for each emotion class. Fig. \ref{fig:music samples} shows a set of music samples generated by our diffusion model, DiT, and JiT. For the detailed demonstrations, please visit our project website: https://diffusiongan.github.io/. 

The symbolic music generated by our model and the other methods were evaluated using a set of metrics. Pitch Class Histogram Entropy (PCHE) \cite{di2021video} assesses the music's quality in tonality by measuring the uncertainty in the distribution of pitch classes and Grooving Pattern Similarity (GPS) \cite{di2021video} quantifies the rhythmicity of music. For these two objective evaluation metrics, the scores closer to real data indicate the higher music quality. The PCHE and GPS results, presented in Table \ref{tab:model_comparison}, suggest our diffusion model is capable of generating samples with higher musical quality than the DiT and JiT baselines using only 10 denoising diffusion steps. Increasing the number of denoising steps to 15 further improves performance. This is much faster than hundreds or thousands of steps required by standard diffusion models such as DiT.  


\begin{table*}
\caption{Subjective evaluation of the generated symbolic music across different criteria.}
\centering
\renewcommand{\arraystretch}{1.5}
\resizebox{0.6\textwidth}{!}
{
\begin{tabular}{cccc}
\specialrule{0.8pt}{0pt}{0pt}
 Model & Humanness$\uparrow$ & Rhythm$\uparrow$ & Overall Preference$\uparrow$             \\ \hline

DiT    &   57.39 $\pm$ 14.22             &   58.21 $\pm$ 14.61                                 &  58.82 $\pm$ 13.62                            \\
JiT    &    65.57 $\pm$ 9.58          &  67.18 $\pm$ 9.59                                  & 70.18 $\pm$ 11.64                               \\
DiffGAN &   69.55 $\pm$ 9.70    & 72.36 $\pm$ 10.46                        & 71.07 $\pm$ 9.86      \\
\specialrule{0.8pt}{0pt}{0pt}
\end{tabular} 
}

\label{tab:subjective_comparison}
\end{table*}



\begin{figure}
\centering
\includegraphics[width=0.5\textwidth]{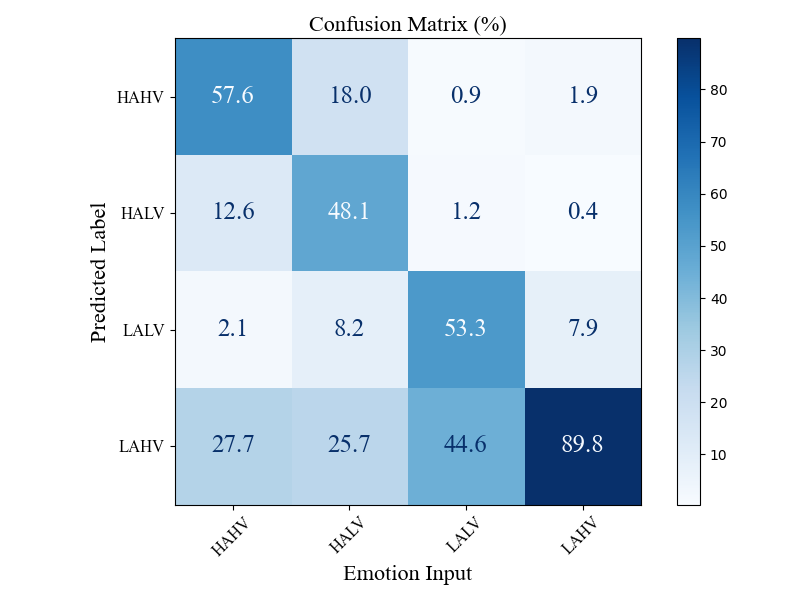}
\caption{\label{fig:heatmap} Accuracy calculated by assessing whether the emotion of our DiffGAN model's generated pieces predicted by the classifier meet the target emotion.}
\end{figure}

In addition, an emotion classifier based on ResNet-152 \cite{he2016deep} was trained to assess whether the generated music's emotions meet the conditions fed to our generative model. The emotion classifier's accuracy on the test set is 0.654. Note that music emotion conditioning is still an open problem. Table \ref{tab:model_comparison} also shows the advantages of our model compared to other methods in terms of emotion accuracy. Our diffusion-based generative model with a denoising step of 10 can produce music with emotions that is consistent with given conditions at an overall accuracy of 0.622. This is a promising result, outperforming the JiT model with the same number of denoising steps and the DiT model with a significantly larger number of steps. Fig. \ref{fig:heatmap} indicates valence is more complex and subtle than arousal and it is more difficult for machine learning models to learn valence as 0.446 of the emotion classifier's outputs are LAHV when the condition fed to our generative model is LALV. This finding is also consistent with those reported in the literature \cite{barthet2013music, hung2021emopia}. 







Overall, our DiffGAN achieves better performance compared to the baselines, as indicated by both the music quality metrics and the emotion classifier.  Instead of generating
samples in one shot, the proposed DiffGAN model produces symbolic music samples by iteratively refining samples through simple denoising steps in which each step is relatively simple to model. We observed that DiffGAN achieves better evaluation results when using 15 denoising steps compared to 5 or 10 steps. This demonstrates the benefits of decomposing the generation process moderately. Moreover, although JiT also allows significantly reduced steps compared to standard diffusion models, our experiments found that its performance deteriorates substantially when the number of denoising steps is excessively reduced to 5, causing it to struggle to generate plausible samples. Our proposed model, however, still achieves promising evaluation results even using a step of 5, demonstrating the advantage of our DiffGAN for efficient sampling.

Complementing the objective evaluations, a subjective listening test was conducted to assess the perceived musical quality of the symbolic music generated by our DiffGAN with 15 steps and the baselines. A total of 15 participants took part in the study. Each model generated one symbolic music sample for each of the four emotion classes, resulting in twelve samples for evaluation across the three models. All samples were presented in a randomized order, with their generating models anonymized. Participants were asked to listen to each sample and provide ratings on a 100-point rating scale across multiple perceptual criteria, including Humanness, Rhythm, and Overall Preference. The results of the listening test are presented in Table \ref{tab:subjective_comparison}. The DiffGAN achieves the highest scores across the three subjective metrics. The Rhythm scores of DiffGAN, JiT, and DiT decrease sequentially, showing a trend consistent with the objective GPS results. This indicates good agreement, as GPS measures the rhythmicity of music.


\section{Conclusion and Future Work}
\label{sec:illust}

We proposed a music generative model that leverages both diffusion models and GANs, enabling fast sampling and controllable generation of symbolic music. Our model produces high-quality samples while taking only 15 diffusion steps to denoise. We use classifier-free guidance to steer the generation towards a target emotion, achieving high overall emotion control accuracy. This contributes to alleviating one of the remaining challenges in machine learning-based music generation, namely music generation controlled by emotions. Currently, our model is limited to the generation of music with four coarse emotion classes. In the future, we plan to explore music generation controlled by continuous arousal and valence values. It is also worth
investigating the use of R3GAN \cite{huang2024gan} which is a modern GAN that could potentially improve DiffGAN for music generation, as it avoids the ad-hoc tricks commonly used in existing GAN models and enhances training stability.

\bibliographystyle{IEEEbib}
\bibliography{icme2026references}

\vspace{12pt}

\end{document}